# Single-shot incoherent imaging with extended and engineered field of view using coded phase apertures

Sai Deepika Sure[1,*], Jawahar Prabhakar Desai[1], and Joseph Rosen[1,2]

[1]School of Electrical and Computer Engineering, Ben-Gurion University of the Negev, P.O. Box 653, Beer-Sheva 8410501, Israel

[2]University of Stuttgart, Institute of Applied Optics, Pfaffenwaldring 9, 70569 Stuttgart, Germany

*sure@post.bgu.ac.il

Abstract

A large field of view of an optical system is needed for many applications, and optical systems with high magnification often suffer from a limited field of view due to the limited size of the camera sensor. This study proposes a novel technique for engineering the field of view of an optical system without compromising the magnification. In the proposed method, an object response pattern is recorded on a camera by introducing a coded phase mask (CPM) in the imaging system. The coded phase mask is a multiplexing of $N$ distinct scattering phases, where $N$-1 represents the number of isolated object areas to be brought within the field of view. Each scattering phase yields a point spread function of a unique sparse dot pattern on the camera. With the introduction of a coded phase mask, the objects' images are brought within the region of the camera sensor, which, without the CPM, would have remained outside the inherent field of view of the system. To reconstruct the original object plane with $N$ objects at their respective locations, the zero-padded object response pattern is deconvolved with the system's zero-padded and shifted point spread function. A simulation study followed by experimental results for $N = 2$ and $N = 3$ is presented in this article.

## 1. Introduction

The field of view is one of the characteristics of an imaging system, which determines the maximum angular width in the object space that can be imaged on the image plane. For a single-lens imaging system focused at infinity, the field of view is limited by the ratio between the sensor size and the lens's focal length. On the other hand, to image an object plane located at a finite distance, the field of view in distance units is given by the ratio of the camera sensor size to the magnification. The use of cameras with a broader sensor size involves an increase in the cost of an imaging system, and it is also less useful for compact imaging systems, where space is another constraint. Extending the field of view (FOV) in imaging systems has been a longstanding challenge, and researchers have explored both optical and computational approaches. Traditional optical solutions, such as fisheye and wide-angle lenses, can capture a broader scene but introduce severe distortions and aberrations at the image periphery, which require complex rectification algorithms for correction[1]. Multiple cameras and panoramic systems[2] offer an alternative by stitching images from multiple viewpoints; however, they increase the system size, complexity of alignment, and cost. Multiple sensor arrays[2] have also been proposed, but they face challenges related to calibration and synchronization. Another approach is to increase the sensor size, which directly enlarges the FOV. However, this solution significantly increases the cost of imaging systems and is unsuitable for compact platforms such as mobile devices or lightweight telescopes.

With the rise of computational imaging, new techniques have been developed to bypass these physical limitations. Image multiplexing and stitching methods[2] can extend the FOV virtually by combining multiple frames; however, they are prone to registration errors, parallax artifacts, and motion sensitivity.

Fourier ptychography[3] has shown that both the resolution and FOV can be enhanced simultaneously by illuminating the sample from multiple angles and applying iterative phase retrieval algorithms[4]. While powerful, this method requires multiple exposures and long reconstruction times, which limit its real-time applicability.

Coded aperture imaging (CAI)[5] represents another important direction, where the FOV and resolution are engineered through optical encoding and computational reconstruction. Early implementations using coded amplitude masks, pioneered by Dicke and Ables for X-ray and astronomical imaging, demonstrated improvements in the signal-to-noise ratio by replacing single apertures with distributed pinholes[6,7]. Later, phase-coded masks (CPMs) became well-used because of their ability to encode more information and improve efficiency when implemented on spatial light modulators (SLMs)[8-10]. In 2016, coded aperture correlation holography (COACH) was introduced as an incoherent digital holography method based on self-interference[8]. Although effective, COACH suffers from tuning problems due to the interference of two beams. To address this, interferenceless coded aperture correlation holography (I-COACH) was developed, eliminating the need for beam splitting and allowing more optical power to reach the detector[9-13]. More recent work has demonstrated compressive I-COACH methods with improved reconstruction quality and power efficiency[14] and I-COACH with interplane crosstalk suppression via deep learning algorithms[15].

To extend both the FOV and the depth of field in imaging, Nakamura et al. proposed computational superposition methods in a special setup of a compound eye[16]. Rai et al. introduced a scattering window technique within the I-COACH framework to address the limited field-of-view inherent to sensor dimensions[17]. By laterally translating a point object and recording multiple peripheral segments of the point spread function (PSF), the segments are stitched into a synthetic hologram nearly nine times wider. This method enables the reconstruction of extended FOV regions via cross-correlation of stitched object response patterns (ORPs) with the stitched PSF. Similar methods for extending the FOV via PSF engineering were proposed in Refs. 18-20. Despite these advances, compact devices such as mobile phone cameras, digital microscopes, and telescopes remain limited by small sensor sizes, typically in the range of 5–7 mm, which restricts the FOV[2,16,17]. Recent reviews of incoherent coded aperture imaging techniques[21,22] emphasize that engineered masks and computational algorithms provide a promising pathway to overcome the FOV limitation by decoupling the FOV from sensor dimensions.

This study introduces a method to expand the field of view of a single optical system without affecting its magnification. The technique involves capturing an ORP on a camera by inserting a CPM into the imaging system. The recorded PSF is related to the CPM transparency through a Fourier transform. The CPM is a multiplexed mask comprising $N$ different scattering phases, with $N$-1 corresponding to the number of separate object regions to be included within the extended FOV. Each scattering phase generates a PSF of a distinct sparse dot pattern on the camera sensor. By incorporating the CPM, objects that normally lie outside the system's inherent FOV are brought into the sensor capture area. To reconstruct the original object plane containing $N$ objects at their respective positions, the zero-padded ORP is deconvolved with the zero-padded and shifted PSF. The optimal sparse pattern is the one that yields a maximum signal-to-noise ratio. The following sections present simulations for $N$=3 and experimental results for the cases of $N$=2 and $N$=3. The manuscript consists of five sections. In the second section, the theoretical description of the proposed method is discussed. Simulation studies and experimental results are presented in the third and fourth sections, respectively. A summary and conclusions are presented in the last section.

## 2. Methodology

The optical imaging system under investigation is shown in Fig. 1, where an object plane is located at a distance $Z_s$ in front of lens $L_1$ of focal length $f_1$, another lens $L_2$ of focal length $f_2$ is located at a distance $Z_h$ from $L_1$, and a camera is located at a distance $Z_o$ behind $L_2$. The imaging condition between the object and image sensor planes is satisfied. The object plane, containing $N = 3$ isolated object areas, and the image recorded in the camera are shown in Figs. 1(a) and 1(b), respectively. Owing to the limited area of the camera sensor, only the central object remains inside the field of view of the camera.

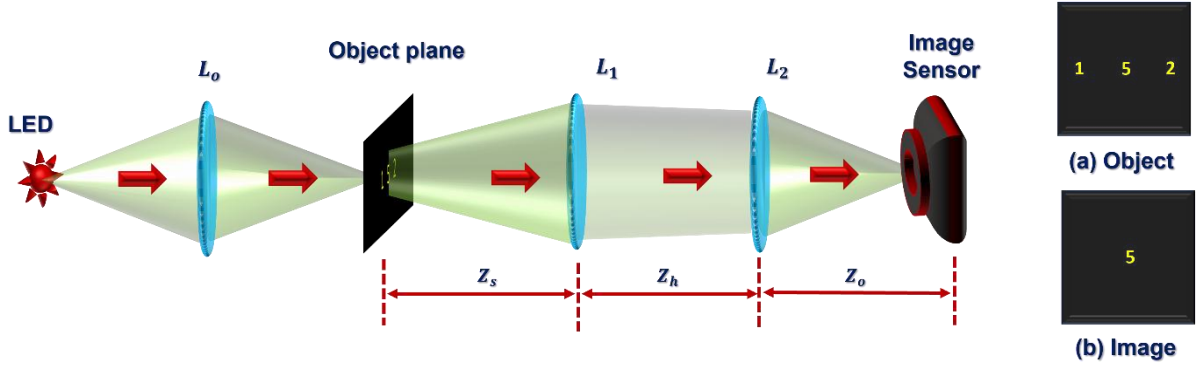

**Fig. 1.** Schematic of the optical system. (a) Object and (b) image recorded by the camera

As shown in Fig. 2, to be able to image the entire object plane, we introduce a CPM in front of lens $L_1$, which is a multiplexing of $N$ scattering masks, where $N$ is the number of object areas; only one of them, the central area, is imaged by the system of Fig. 1. The design shown in Fig. 2 ensures that the camera plane is in a Fourier relation with the entire CPM transparency when the PSF is recorded by the camera. The Fourier relation is achieved because of the imaging condition between the object plane and the camera plane, and its mathematical meaning is that the intensity in the camera plane is relative to the magnitude square of the scaled 2D Fourier transform of the CPM transparency [the mathematical description is given in Eq. (2)][11]. As mentioned, the CPM is a multiplexing of $N$ scattering phases, where each scattering phase is assigned to one of the $N$ object areas. When there is a point in the object plane, every scattering phase is designed to yield a unique intensity pattern of $K$ sparse dots in its Fourier transform, with a weak correlation with the other dot patterns. Additionally, the $n^{th}$ scattering phase is generated such that although the $n^{th}$ object point is out of the system's FOV, its response is obtained inside the camera sensing area. Therefore, in the presence of only the $n^{th}$ object point and the $n^{th}$ scattering phase, $K$ replicas of the $n^{th}$ object appear within the field of view of the camera. The $n^{th}$ scattering phase also modulates the light coming from other $N-1$ objects in the object plane, creating $K$ replicas for each of them but outside the field of view. The Fourier relation between the camera plane and the CPM enables us to use the modified Gerchberg–Saxton algorithm (GSA) to synthesize the scattering phase function, as shown in Fig. 3.

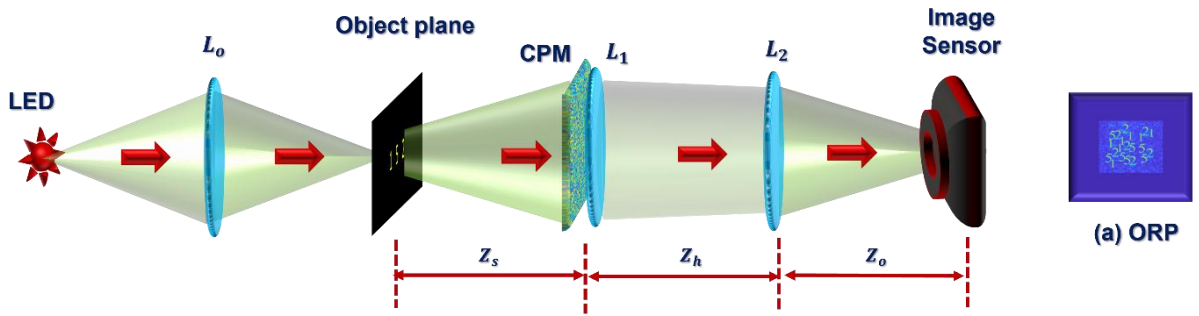

**Fig. 2.** Optical system for engineering the field of view: (a) Object response pattern (ORP).

A uniform matrix as a magnitude and a random matrix in the range [0, 2π] as an initial phase are taken as an input at the CPM plane of the GSA. The magnitude of the Fourier transform of the CPM matrix is replaced by the desired sparse dot pattern, while the phase is retained and introduced at the camera plane for calculating the inverse Fourier transform. At the end of the predefined number of iterations, the phase part at the CPM plane is used as the scattering phase function of the system.

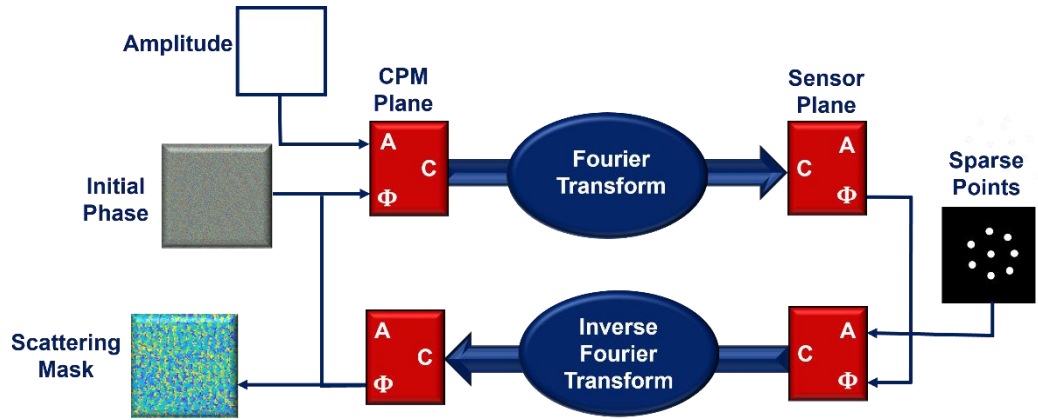

**Fig. 3.** Schematic of the modified Gerchberg–Saxton algorithm for generating the $n^{th}$ scattering mask.

$N$ such coded phase masks are synthesized via the above method. The technique used for multiplexing $N = 3$ CPMs is depicted in Fig. 4. CPMs 1-3 synthesized via the above GSA method are shown along row (a) in Fig. 4. CPM2 and CPM3 are designed for the horizontal peripheral areas, and hence a high-frequency grating is embedded into their phase to guarantee that the response of any peripheral point appears in the center of the camera plane. Next, a matrix of similar size to the SLM matrix, following a Gaussian distribution in the range [0, 1] is divided into 3 sections. Sections 1-3 are groups of pixel values in the ranges [0, 0.33], (0.33, 0.66], and (0.66, 1], respectively, as shown along row (b) in Fig. 4 from left to right. The group of pixels in Section $n$ is used to display only the $n^{th}$ CPM. For that, CPMs 1-3 shown along row (a) are multiplied by sections 1-3 of Fig. 4(b). The resulting masks are shown in Fig. 4(c). The sum of $N = 3$ masks in row (c) yields multiplexing of CPMs, as shown in Fig. 4(d). The phase of the diffractive lens $L_1$ is added to this CPM to obtain a phase mask, as shown in Fig. 4(e), that is displayed on the SLM and together with lens $L_2$ of Fig. 2 satisfies the imaging condition between object and camera planes. The introduction of the CPM in the system results in the formation of $K$ replicas on the camera for each of the $N$ objects.

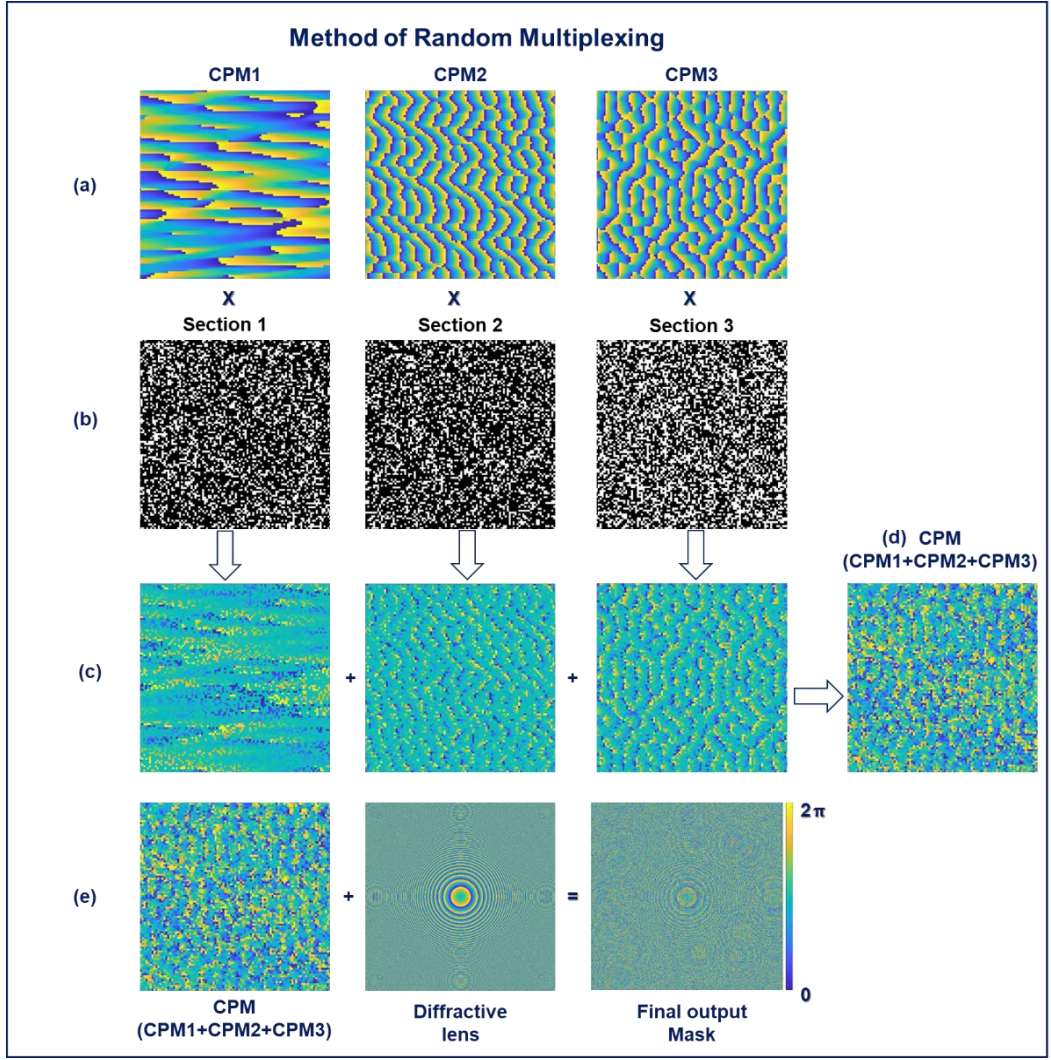

**Fig. 4.** Multiplexing of scattering phases: (a) the phase of three distinct CPMs and (b) binary-valued [1,0] random masks in which the product of any two different masks is a zero matrix. (c) Product of coded phases and random sections. (d) The multiplexed CPM phase is obtained as the sum of the three phases. (e) The final phase of the aperture mask is obtained as the sum of the phases of the CPM and the diffractive lens.

In the following mathematical analysis of the proposed method, each object is a collection of point sources. Considering a point source at the center of the $n^{th}$ object area, with coordinates $(\bar{r}_n, z_s) = (x_n, y_n, z_s)$ and with an amplitude $a_n$, here $\bar{r}_n$ denotes the transverse location vector to the center of the $n^{th}$ object area. The complex amplitude incident on the CPM is given as $a_n L(\bar{r}_n/z_s)Q(1/z_s)$, where $L$ and $Q$ represent linear and quadratic phase functions, respectively, given by $L(\bar{r}_n/z) = \exp[i2\pi(\lambda z)^{-1} \cdot (x_n x + y_n y)]$ and $Q(1/z) = \exp[i\pi(z\lambda)^{-1} \cdot (x^2 + y^2)]$, where $\lambda$ is the wavelength of light and $z$ is the axial distance. The complex amplitude modulated by the $n^{th}$ scattering phase attached to the diffractive lens $L_1$ is given by $a_n L(\bar{r}_n/z_s)Q(1/z_s)\exp[i\Phi_n]L(-\bar{r}_n/z_s)Q(-1/f_1)$, where $\Phi_n$ is the $n^{th}$ scattering phase of the CPM calculated via the modified GSA, $L(-\bar{r}_n/z_s)$ is the linear phase attached to the $n^{th}$ CPM to deflect the $n^{th}$ PSF into the center of the camera plane, and $f_1$ is the focal length of the diffractive lens $L_1$. The field incident on lens $L_2$ is given by $a_n Q(1/z_s)\exp[i\Phi_n]Q(-1/f_1) * Q(1/z_h)$, where $*$ denotes the convolution. After passing through lens $L_2$, the complex amplitude is given by $\{a_n Q(1/z_s)\exp[i\Phi_n]Q(-1/f_1)\} * Q(1/z_h)Q(-1/f_2)$, where $f_2$ is the focal length of lens $L_2$. The intensity obtained at the sensor plane is given by Fresnel propagation along the distance $z_o$, as follows,

$$PSF_n(\bar{r}_o; \bar{r}_n) = \left| \left[ \left\{ a_n Q\left(\frac{1}{z_s}\right) \exp[i\Phi_n] Q\left(\frac{-1}{f_1}\right) * Q\left(\frac{1}{z_h}\right) \right\} Q\left(\frac{-1}{f_2}\right) \right] * Q\left(\frac{1}{z_o}\right) \right|^2, \quad (1)$$

where $\bar{r}_o = (u, v)$ is the position vector on the camera plane. In the case of a single point in the input plane that is imaged to the camera plane by lenses $L_1$ and $L_2$, the CPM and camera planes are in a Fourier relation; therefore, the intensity response to a point located in the object plane at $(\bar{r}_n, z_s)$ is given by the following expression:

$$PSF_n(\bar{r}_o; \bar{r}_n) = \left| v\left[\frac{1}{\lambda z_L}\right] \mathcal{F}\left\{ a_n L\left(\frac{\bar{r}_n}{z_s}\right) L\left(\frac{-\bar{r}_n}{z_s}\right) \exp[i\Phi_n] \right\} \right|^2 = PSF_n(\bar{r}_o + M\bar{r}_n; [0,0]), \quad (2)$$

where $v[\cdot]$ is the scaling operator such that $v[\alpha]f(x) = f(\alpha x)$, $\mathcal{F}\{\}$ is a 2D Fourier transform operator, $M$ is the system's transverse magnification, and $z_L = (z_h f_2 - z_h z_o + z_o f_2)/f_2$. Eq. (2) indicates that when the input point is at the center of the $n^{th}$ object area, the PSF is obtained in the center of the camera plane. The light coming from a point source at the $n^{th}$ object area and modulated by the remaining $N$-1 scattering phases creates $K(N-1)$ dots outside the field of view of the camera; hence, the intensity on the camera $I_c$ remains the same as in Eq. (2), as in the following equation,

$$I_c(\bar{r}_o) = \left| v\left[\frac{1}{\lambda z_L}\right] \mathcal{F}\left\{ L\left(\frac{\bar{r}_n}{z_s}\right) \sum_{m=1}^{N} a_m L\left(\frac{-\bar{r}_m}{z_s}\right) \exp[i\Phi_m] \right\} \right|^2 = PSF_n(\bar{r}_o; \bar{r}_n). \quad (3)$$

For $N$ point sources located at $r_n = r_1, r_2, \ldots, r_N$, the total PSF is given by,

$$PSF = \sum_{n=1}^{N} PSF_n. \quad (4)$$

Any 2D object is an ensemble of a shifted sum of $l$ delta functions; therefore, the intensity created by an object at the $n^{th}$ area, at the camera coordinates, is given as follows,

$$I_n(\bar{r}_o) = \sum_l b_{n,l} \delta[\bar{r}_o - M(\bar{r}_n + \bar{r}_l)]. \quad (5)$$

The intensity response for the $n^{th}$ object area, as recorded on the camera, is obtained as a convolution between the object intensity and the $PSF_n$, as follows,

$$ORP_n = \sum_l b_{n,l} \delta[\bar{r}_o - M(\bar{r}_n + \bar{r}_l)] * PSF_n = \sum_l b_{n,l} PSF_n(\bar{r}_o - M\bar{r}_l). \quad (6)$$

Similarly, the total $ORP$ is given by,

$$ORP = \sum_{n=1}^{N} \sum_l b_{n,l} PSF_n(\bar{r}_o - M\bar{r}_l). \quad (7)$$

For reconstruction of the image of the object plane, $ORP$ is zero-padded to the size of the magnified object plane, and the various PSFs are positioned at the corresponding regions in the big matrix with the same size. The reconstructed image of the entire object plane is,

$$I_{REC} = ORP \circledast \sum_{n=1}^{N} PSF_n(\bar{r}_o - M\bar{r}_n), \quad (8)$$

where $\circledast$ indicates deconvolution. Note that reconstruction with part of the group of PSFs yields an image of only part of the entire object plane. Furthermore, changing the locations of the various PSFs in the reconstruction matrix yields an image of the object plane, but in a different order of the various regions. In that sense, the proposed technique is FOV engineering.

## 3. Simulation results

The optical system shown in Fig. 2 is simulated in MATLAB. The size of the object plane is 1080 × 1080 pixels, whereas the size of the camera sensor is limited to 500 × 500 pixels, with a pixel size of 8 µm. The distances $z_s = z_h = 20$ cm, and $z_o = 10$ cm. The focal length of lens $L_1$ is $f_1 = 20$ cm, and for lens $L_2$, it is $f_2 = 10$ cm. The CPMs, ORPs, reconstructed images, and zoomed-in regions around objects 1, 2, and 3 in the reconstructed images are shown along columns 1-4, respectively, from top to bottom for $K = 3$-12 in Fig. 5. The objects lying in the object plane and the image recorded in the camera are shown in the insets of Fig. 5 labeled (a) and (b), respectively. To mimic the experimental conditions, Poisson noise is added to both the PSF and the ORP via the built-in MATLAB function imnoise(·,"poisson"). Reconstruction is carried out with the Wiener deconvolution[23] technique via the built-in function deconvwnr(ORP,PSF,NSR), where NSR represents the noise-to-signal ratio. The parameter signal-to-noise ratio (SNR) is used as a metric to determine the optimal number of dots in a sparse pattern, and the plot of the SNR versus $K$ is shown in Fig. 6. According to Fig. 6, the highest SNR is obtained for a PSF of 11 dots.

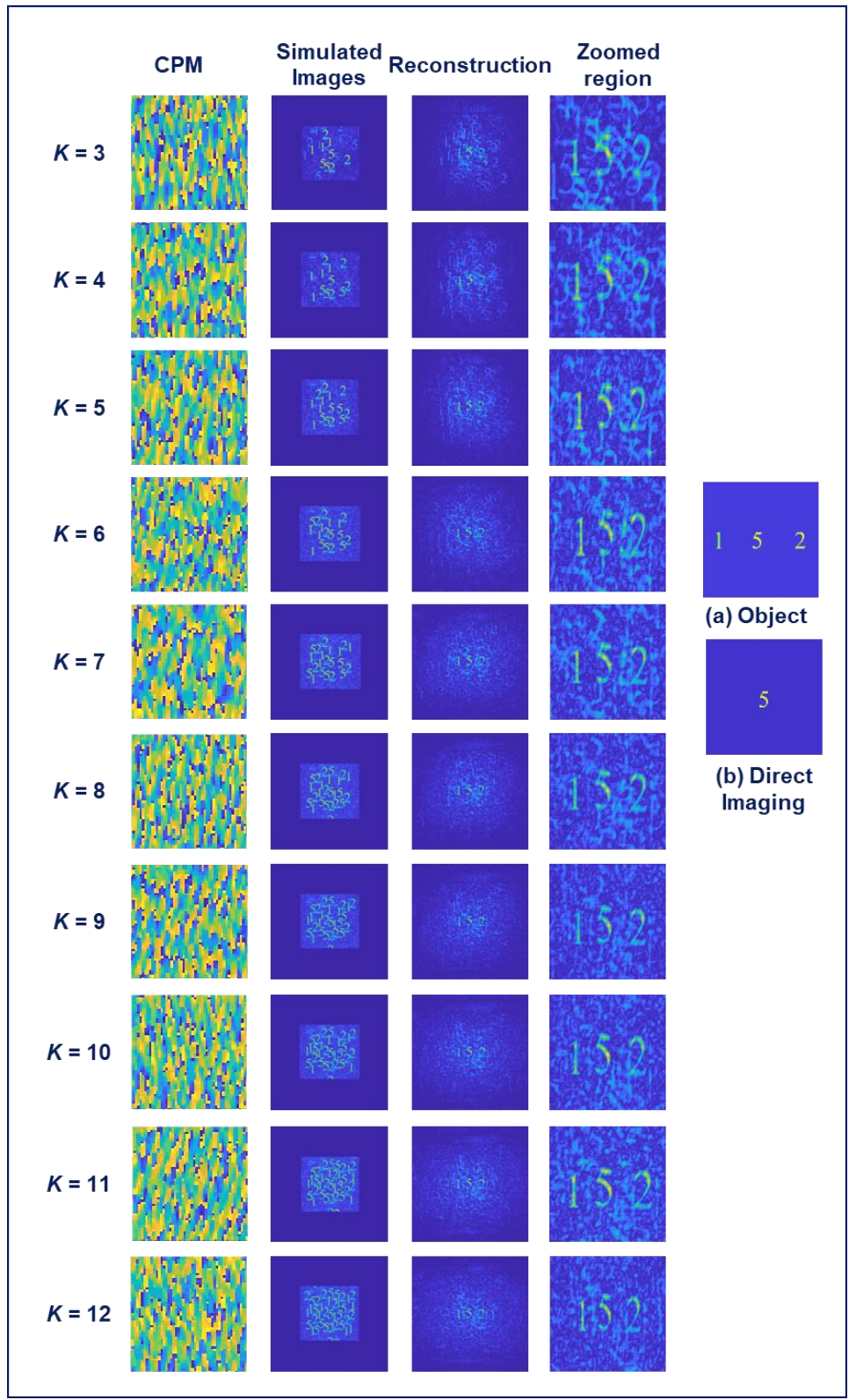

**Fig. 5**. Simulation results for the number of sparse dots *K*=3-12 from top to bottom. Each row from left to right shows part of the CPM phase, ORP, reconstruction via Wiener deconvolution, and zoomed-in images of the three objects. The insets are (a) the object plane and (b) the image recorded in the camera for the direct imaging system of Fig. 1.

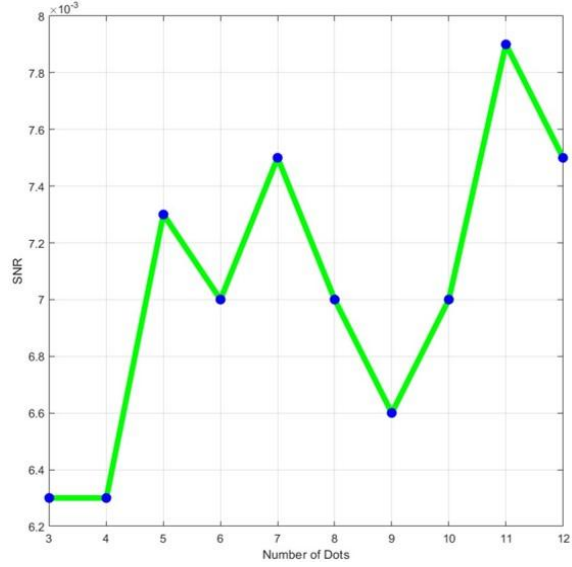

**Fig. 6.** Plot of the *SNR* versus *K* in the simulation.

## 4. Experiments

The experiment of this method is carried out under two different conditions. In the first case, two different yet identical channels are used to place two objects at different transverse positions $\bar{r}_n$ in the object plane at an axial distance $z_s$. In this case, the field of view of the system is limited by the sensor area of the camera. In the second case, the field of view is further reduced, and the utility of this study is verified. An experiment is carried out for the same magnification as before, where three objects are located at different transverse positions $\bar{r}_n$, placed in one channel at an axial distance $z_s$, and illuminated with a single light-emitting diode (LED), which creates a narrow illumination spot on the object plane. Closely lying objects are chosen from the United States Air Force (USAF) target to maintain uniform illumination of each object. These conditions result in the appearance of all three objects within the sensor area of the camera. Therefore, the field of view of the system is reduced computationally, where the ORP recorded in the camera is masked with a binary mask matrix [1,0] in the computer. This technique limits the field of view irrespective of the physical size of the camera sensor. This technique demonstrates that the method developed in this study is useful for engineering the FOVs of systems with relatively small sensor sizes.

*4.1 Engineering Field of View with two channels:*

A schematic of the first experimental technique is shown in Fig. 7, where two identical channels are used for illuminating two objects. Element 3 in Group 5 and Element 2 in Group 4 of the positive USAF 1951target are used as object 1 [Fig. 7 (a)] in Channel 1 and object 2 [Fig. 7 (b)] in Channel 2, respectively. Both objects are critically illuminated with incoherent LEDs (Thorlabs LED635L, 170 mW, λ = 635 nm, Δλ = 15 nm) and two identical lenses, $L_A$ and $L_B$, of focal lengths $f_O = 10$ cm in their respective channels. The LEDs are kept before the $L_A$ and $L_B$ at $z_1 = 20$ cm, whereas the objects are beyond the $L_A$ and $L_B$ at a distance of $z_2 = 20$ cm. The light transmitted through object 2 is directed toward the SLM with a BS1 beam splitter. An SLM (Holoeye PLUTO, 1920 × 1080 pixels, 8 μm pixel pitch, phase-only modulation) is used to display a phase mask. In the experiment, the glassy lens $L_1$ is replaced with a diffractive lens (DL) with a focal length of $f_1 = 20$ cm, which is combined with the CPM and displayed on the SLM, as shown in Fig. 7(c). The SLM is a polarization-sensitive device that can modulate only the component of light directed along its active axis. A polarizer is used to eliminate the light orthogonal to the active axis before it is incident on the SLM. The SLM is placed at $z_s = 20$ cm from both objects. The light reflected from the SLM is modulated according to the CPM, combined with a diffractive lens displayed on it. The reflected light is then directed toward lens $L_2$ of focal length $f_2 = 10$ cm, placed at $z_h = 20$ cm from the SLM, using another beam splitter $BS_2$. From lens $L_2$, the light is incident on the camera (Thorlabs 8051-M-USB, 3296×2472 pixels, 5.5 µm pixel pitch) placed at $z_o = 10$ cm. The image recorded by direct imaging of two lenses without the CPM is shown in Fig. 7(d), where object 1, located at the center of the object plane, remains in the field of view. In contrast, object 2, which is placed horizontally to the left, is outside the field of view of the camera sensor. In the presence of the CPM, which is the multiplexing of $N = 2$ scattering phases, object 2 should be imaged owing to the extension of the field of view.

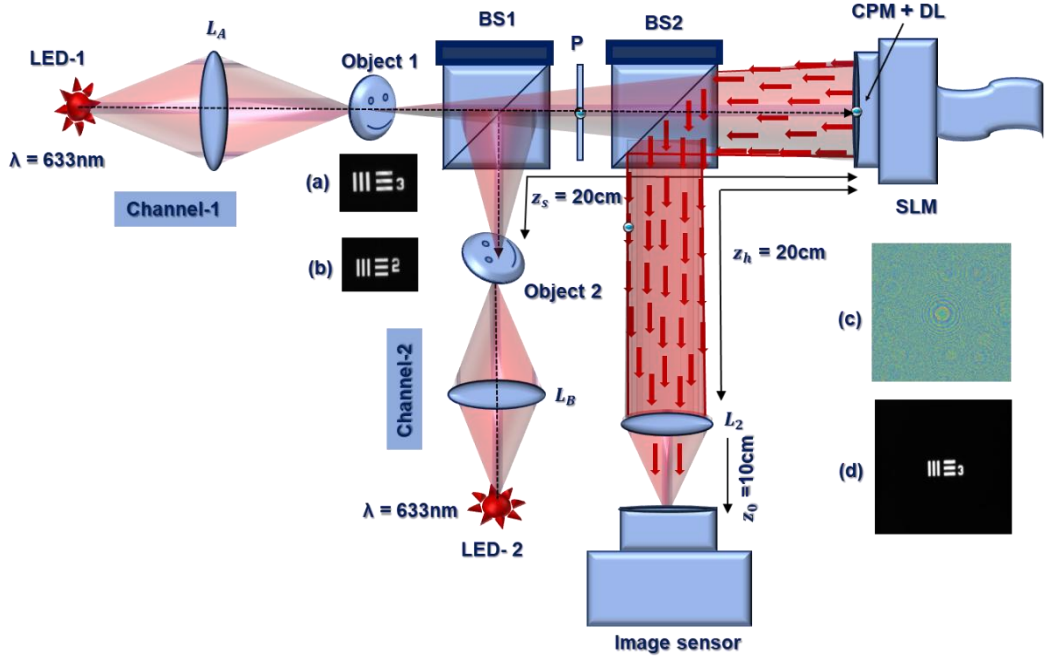

**Fig. 7.** Experimental setup with two channels used for illuminating two objects. Insets: (a) object 1, (b) object 2, (c) CPM + DL displayed on the SLM, and (d) image recorded in the direct imaging system, where LED-1 and LED-2 are light–emitting diodes; $L_A$ and $L_B$ – illumination lenses; P – polarizer; BS1 and BS2 – beam splitters; CPM – coded phase mask; DL – diffractive lens; SLM – spatial light modulator; ⊖– Polarization orientation directed along the SLM plane is shown in Fig. 1.

The experiment is carried out for varying numbers of sparse dots from 3-11. The CPM, ORP, reconstruction, and zoomed area of object 1, outlined in red, and object 2, outlined in green, in the combined reconstruction are shown along columns 1-5 for $K=3$-12 from top to bottom in Fig. 8. To avoid overlapping objects 1 and 2 in the reconstruction, $PSF_2$ is shifted by 650 pixels to the left. The ORP is deconvolved with the shifted superposition of $PSF_1$ and $PSF_2$. The region outlined with red and green colors contains reconstructions of objects 1 and 2, respectively; hence, it is considered a signal, whereas the region outside the boxes corresponds to noise. The average SNR is used as a figure of merit to determine the optimal number of dots.

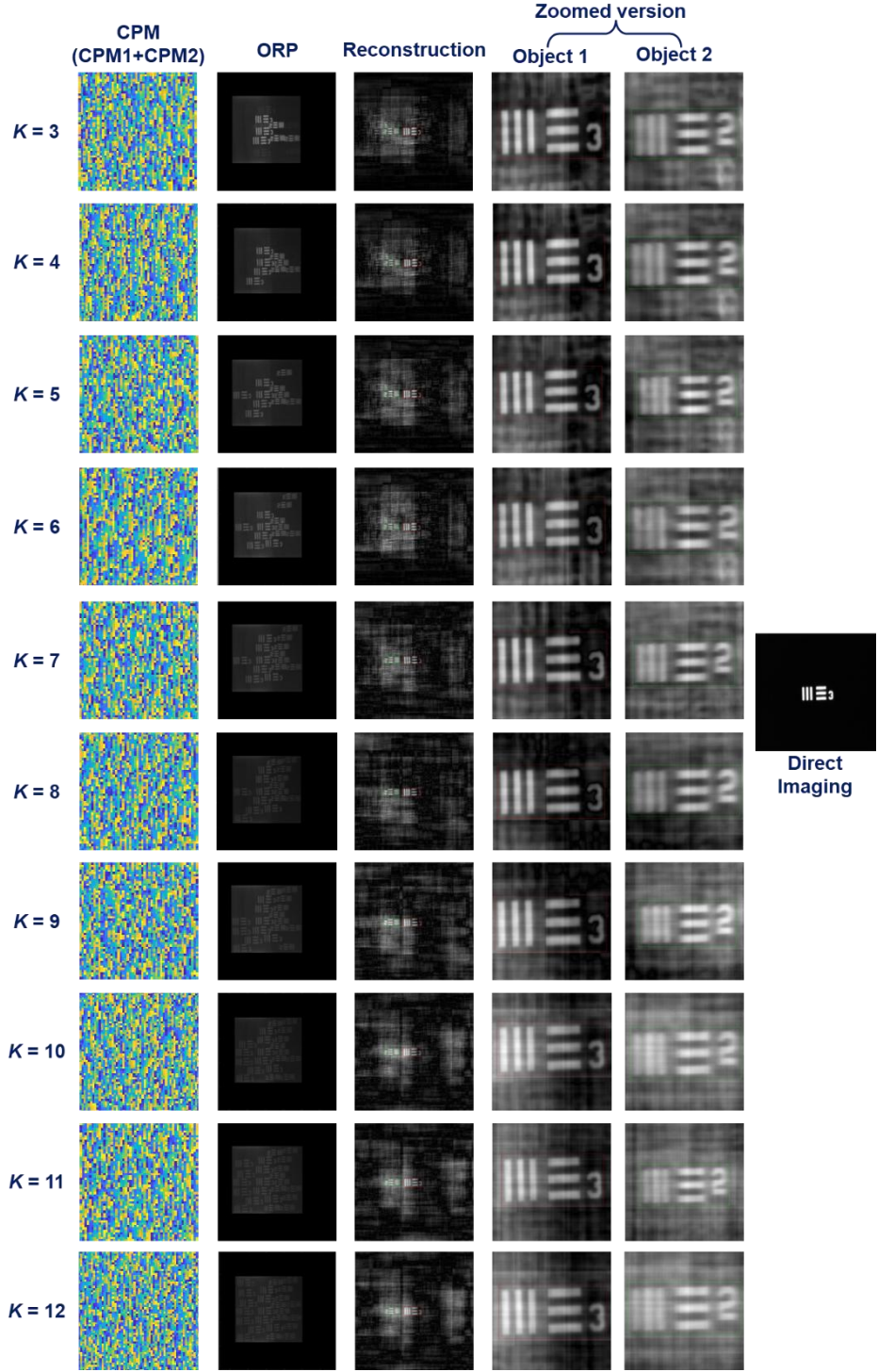

**Fig. 8.** Experimental results for the number of sparse dots $K$ = 3-12 from top to bottom. Along rows (from left to right), part of the CPM phase, ORP, reconstruction, zoomed region around object 1, and object 2 taken from the reconstruction, and direct imaging in the inset.

Different metrics are used to determine the optimal number of dots in the sparse pattern. The signal-to-noise ratio (SNR) is calculated with the expression below.

$$SNR = \frac{\sum_{n=1}^{N} sum\ of\ pixel\ values\ inside\ outlined\ region/area\ of\ outlined\ region}{\sum sum\ of\ pixels\ values\ outside\ outlined\ region\ /area\ outside\ outlined\ region} \qquad (9)$$

The mean squared error (MSE) is calculated between the reconstructed image and the direct image via the MATLAB built-in function 'immse'. Another metric, the structural similarity index measure (SSIM), is used to determine the optimal number of dots. The SSIM is calculated via the MATLAB built-in function 'ssim' for the reconstructed image and the reference image recorded via direct imaging. The graphs of the SNR, SSIM, and MSE versus the number of dots are shown in Fig. 9(a), 9(b), and 9(c), respectively. The system with a PSF of $K=8$ dots has the maximum SSIM, minimum MSE, and second-best SNR; hence, $K=8$ is the optimal value of this experiment.

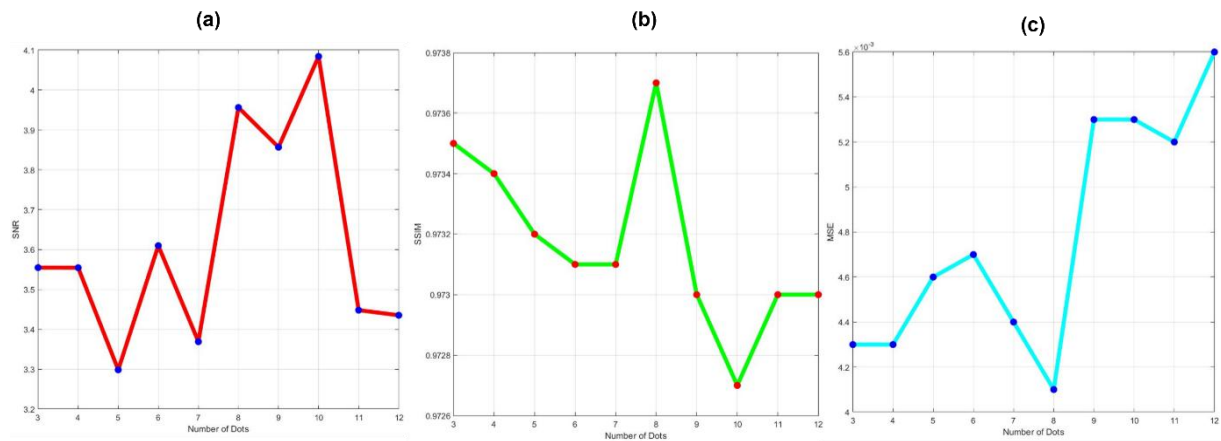

**Fig. 9.** Plots of (a) the SNR versus $K$, (b) the SSIM versus $K$, and (c) the MSE versus $K$ in the first experiment.

*4.2 Engineering Field of View with a single channel:*

A schematic of the next experimental setup is shown in Fig. 10, where a single channel is used to illuminate three different objects of different sizes in the USAF target. In this experiment, incoherent light from an LED illuminates an object plane consisting of three objects. Element 5 in Group 3, Element 4 in Group 2, and Element 6 in Group 1 of the positive chart USAF 1951 are used as objects 1, 2, and 3, respectively, which are placed at a distance $z_s = 20$ cm from the SLM. All the objects are critically illuminated by an LED and lens $L_O$. The light transmitted through objects is passed through a polarizer, which is directed along the active axis of the SLM to block unmodulated light. The CPM, which is a multiplexing of three scattering phases combined with a diffractive lens, is displayed on the SLM. The light reflected from the SLM is directed toward lens $L_2$ of focal length $f_2 = 10$ cm located at $z_h = 20$ cm from the SLM with a beam splitter BS. The light refracted from lens $L_2$ is incident on the camera, where the ORP is recorded.

An image recorded via direct imaging is shown in Fig. 10(b), which illustrates the respective locations of objects 1-3. The field of view is limited on the computer by padding with zero outside a rectangular area of 566 × 1183 pixels. The image with a limited field of view is shown in Fig. 10(c), demonstrating that only object 1 is within the field of view, whereas objects 2 and 3 are outside the FOV. To avoid overlap between the reconstructed objects, ORP, $PSF_1$, $PSF_2$, and $PSF_3$ are zero-padded to form a larger field of size 7416 × 9888 pixels (i.e., 3 × 2472 in height and 3 × 3296 in width). Each PSF is then positioned within this extended grid by applying specific pixel shifts relative to the center location. $PSF_1$ is placed at the center of the padded field. $PSF_2$ is shifted horizontally 2472 pixels to the left relative to the center. $PSF_3$ is shifted horizontally 2472 pixels to the right of the center. Zero-padded ORP is deconvolved with the zero-padded and shifted sum of $PSF_1$, $PSF_2$, and $PSF_3$. The regions outlined with blue, red, and green colors contain reconstructions of objects 1, 2, and 3, respectively; hence, they are considered the desired signals, whereas the region outside the boxes corresponds to noise. The average SNR is used as a figure of merit to determine the optimal number of dots. The CPM, ORP, reconstruction, and zoomed region around object 1, object 2, and object 3 in the combined reconstruction are shown along columns 1- 6 by varying the number of sparse dots ($K$) from 3 to 7 from top to bottom in Fig. 11.

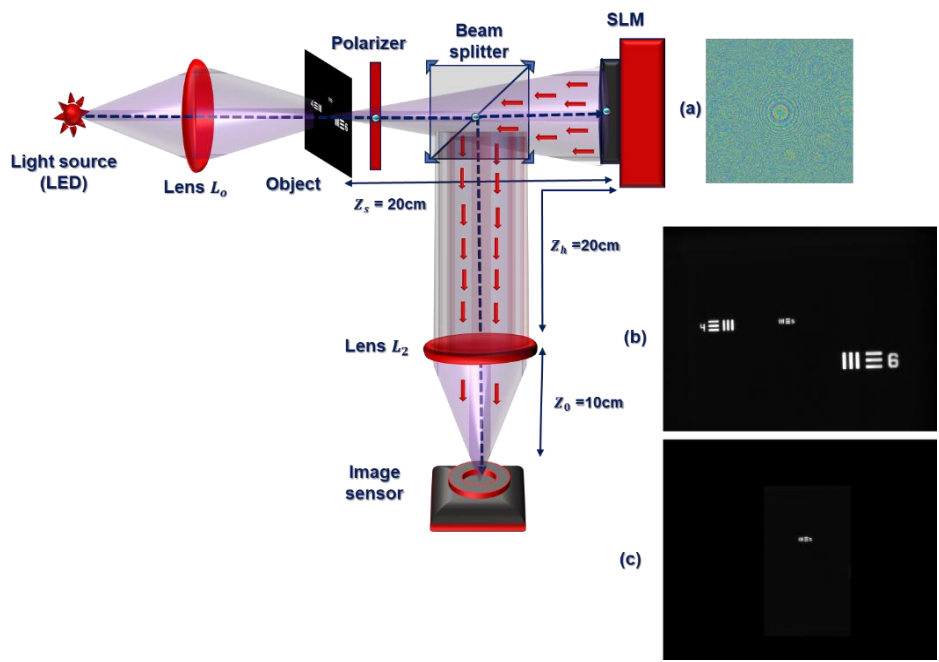

**Fig. 10.** Experimental setup of I-COACH for the engineering field of view. (a) CPM + DL displayed on the SLM, (b) image recorded in the direct imaging system, and (c) direct image after limiting the field of view of the system on the computer.

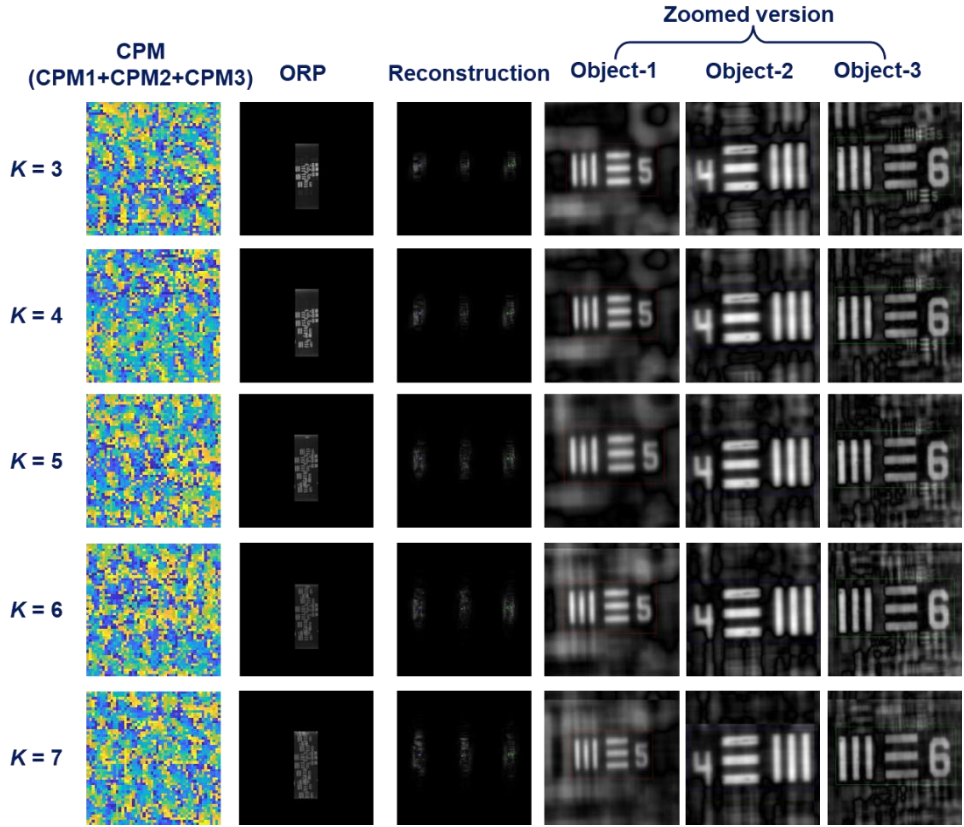

**Fig. 11.** Experimental results for *K* = 3-7 from top to bottom. In the first column, part of the CPM phase (CPM1+CPM2 +CPM3); in the second column, ORP; in the third column, the reconstruction results; and in the fourth, fifth, and sixth columns, zoomed regions from the reconstruction image around objects 1, 2, and 3, respectively.

The graphs of the SNR, SSIM, and MSE versus the number of dots are shown in Fig. 12(a), 12(b), and 12(c), respectively. The case of *K*=7 yields the best figures of merit; therefore, it is the optimal CPM of this experiment.

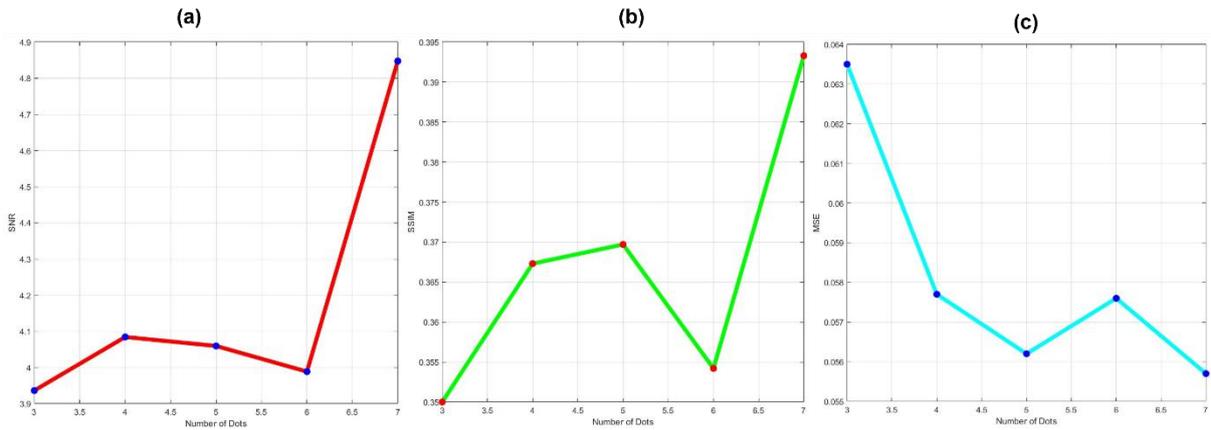

**Fig. 12.** Plot of (a) SNR versus *K*, (b) SSIM versus *K*, and (c) MSE versus *K* in the second experiment.

## 5. Discussion and conclusion

We have demonstrated single-shot incoherent imaging capabilities for engineering the field of view with a phase-coded aperture. In this context, the light scattered from a point object is modulated by the

CPM, which is a multiplexing of *N* different scattering phase masks attached to the diffractive lens and is directed to the sensor plane by the lens $L_2$. The camera records a distinct, sparse dot pattern for every scattering phase when the system is illuminated by a point object. With the help of a phase mask, the objects inside the field of view that are lost due to the limitation of the system's field of view are recovered. The final image of the object is reconstructed via Wiener deconvolution by deconvolving the zero-padded object with the shifted PSFs of the system. SNR, SSIM, and MSE are the metrics taken as comparison criteria to determine the optimal number of sparse dots used in our system. The experiments are carried out for two different setups, demonstrating two different scenarios of the proposed method of extending the FOV. The single-shot imaging capabilities of multiple planes of the object are verified experimentally by postprocessing the recorded ORP. The resulting intensity pattern for *K* = 3–7 dots with a single channel is shown in Fig. 11, where the maximum SNR is at *K*=7, with the lowest MSE and the highest SSIM, as is shown in Fig. 12. In this scenario, we conclude that *K*=7 is the optimal value since it yields three optimal figures of merit.

The extended FOV image in the present method is obtained as a deconvolution between the overall ORP and PSF, each of which is a sum of *N* PSFs and ORPs. Such deconvolution creates a background noise around the image. This noise has two sources: 1. The main source of noise is cross-deconvolution between $ORP_m$ and $PSF_n$, where $m \neq n$. The cross-deconvolution noise can be eliminated by changing the multiplexing method of CPMs on the SLM. In a continued study, which will be reported soon, we consider the issue of CPM multiplexing and show that time multiplexing can eliminate the cross-deconvolution noise, but at the cost of lower time resolution and slowness of the capturing process. 2. The other noise source is the side lobes of the intra-deconvolution between $ORP_n$ and $PSF_n$. This noise is usually lower than the cross-deconvolution noise, but should be eliminated if possible. Our team is in the process of looking for methods to reduce this noise, as well. The other sources of noise are negligible in comparison to the above-mentioned two sources, and hence there is no point in considering them before the main noise is taken care of.

Compared with existing field-of-view extension techniques, the proposed CPM-based multiplexing approach offers a distinct balance of compactness, efficiency, and single-shot operation. Conventional FOV extension techniques, such as wide-angle or fisheye optics and multi-camera panoramas, increase coverage at the cost of peripheral distortion, bulky hardware, or high system complexity, whereas computational methods such as image stitching and Fourier ptychography require multiple exposures and iterative reconstruction, limiting them to static scenes and making them computationally demanding. PSF engineering approaches, such as scattering window techniques and compound-eye designs, can significantly extend the FOV but typically rely on mechanical scanning or multiple measurements. In contrast, the proposed CPM-based multiplexing method enables multi-region FOV extension in a single shot without mechanical scanning while preserving the native magnification and resolution of the imaging system. Reconstruction is efficient and non-iterative, relying on a single deconvolution between the zero-padded ORP and the corresponding PSF. The primary limitation is that the number of multiplexed regions is constrained by the system aperture, leading to an SNR trade-off as the multiplexing order increases.

The proposed concept sets a way to see all objects from multiple areas that are out of the FOV in direct imaging. This method enables one to reconstruct any permutations of the *N* areas just by reconstructing any possible combination of the various PSFs. Moreover, the locations of the various object areas in the reconstructed image can be determined arbitrarily by shifting the various PSFs to arbitrary positions. This project is an additional example of extending the imaging limits by I-COACH with multiplexed CPMs. The first published example in this direction is the 3D I-COACH proposed in Ref. 24, in which the various imaging domains have been subspaces along the *z*-axis of the object space. This family of imaging methods suffers from high noise levels due to the attempt to multiplex more than one CPM in the same single aperture. As mentioned above, to reduce the noise from the reconstructed images, different methods of CPM multiplexing have already been investigated by our group, and the results are expected to be reported soon.

**Funding**

The authors are grateful for funding from the Israel Science Foundation (ISF) Grant No. 3306/25 for supporting this research project.

**Data availability statement**

The datasets used and/or analysed during the current study available from the corresponding author on reasonable request.